\documentstyle[12pt]{article}
\begin{document}

\rightline{FTUV/93-21}
\rightline{IFIC/93-09}

\vspace{0.5cm}

\begin{center}
{\Large \bf COMPLEX ANALYTIC REALIZATIONS \\
\vspace{0.3cm}
FOR QUANTUM ALGEBRAS}\footnote{Supported by DGICYT, Spain.

*azcarrag @ evalvx.ific.uv.es.

**ellinas @ evalvx.ific.uv.es.}
\end{center}

\vspace{3cm}

\begin{center}
{\large \bf J.A. de Azc\'arraga* and D. Ellinas**}
\end{center}

\begin{center}
{\it Departamento de F\'{\i}sica Te\'orica and IFIC,\\
Centro Mixto Universidad de Valencia - CSIC,\\
E-46100 Burjasot, Valencia, Spain}
\end{center}

\vspace{3cm}

\begin{abstract}
A method for obtaining complex analytic realizations for a class of
deformed algebras based on their respective deformation mappings and
their ordinary coherent states is introduced. Explicit results of such
realizations are provided for the cases of the $q$-oscillators
($q$-Weyl-Heisenberg
algebra) and for the
$su_{q}(2)$ and $su_{q}(1,1)$ algebras and their co-products.
They are given in terms of a series in powers of ordinary derivative operators
which act on the Bargmann-Hilbert space of functions endowed with the
usual integration measure.
In the $q\rightarrow 1$ limit these realizations reduce to the usual
analytic Bargmann realizations for the three algebras.
\end{abstract}

\newpage

\baselineskip 0.7cm

\noindent
{\Large \bf I.-Introduction}

The representation theory of quantum algebras and groups \cite{1,2,3,4},
constitutes an open field of research. Some of the features specific to
quantum algebras, for example those appearing when the deformation
parameter $q$ becomes a root of unity, have been studied
already \cite{5,6}. It will be shown here for some of the simplest types
of deformed algebras that also for $q$ real there are some interesting
realizations which can elucidate the relation between deformation and
non-linearity.

We will confine our scope in this paper to three of the simplest quantum
algebras i.e., the so called q-oscillator and the $su_{q}(2)$
and $su_{q}(1,1)$ algebras  \cite{7,8,9,10,11}.
We will seek complex analytic realizations of the above algebras which,
unlike the ones existing in the literature which are based on the so called
$q$-coherent states \cite{12,13,14} and involve $q$-deformed (Jackson)
derivatives
\cite{13,15}, will be given instead in terms of a series of higher powers of
ordinary derivatives. The representation spaces of these deformed realizations
will be the ordinary Hilbert spaces of square-integrable analytic functions
$L^{2} ( \frac{G}{H}, d \mu (\zeta))$, built on the corresponding cosets
of the non-deformed Lie groups, i.e $G/H = \frac{Weyl-Heisenberg}{U(1)},
\frac{SU(2)}{U(1)}$ and $\frac{SU(1,1)}{U(1)}$. The invariant measure of
integration $d \mu (\zeta)$, the so-called Bargmann measure, will be
explicitly given below for each case. One feature of the obtained
realizations of the quantum algebra generators is that they constitute a
deformation of the ordinary Lie algebra generators in the
sense
that they involve a series in powers of ordinary derivatives
(the coefficients of which depend on the deformation parameter
$q$) that reproduces in the  'classical'
$q\rightarrow 1$ limit the Lie algebra vector field
generators.

The outline the paper is as follows: in Section II, the required coherent
states
(CS) formulae for each of the non-deformed Weyl-Heisenberg ($wu$), $su(2)$ and
$su(1,1)$ algebra are given. Also the deformation mappings relating the
quantum versions of the above algebras to the generators of the respective
non-deformed ones will be provided, as they will be important in the next
Section for the analytic realizations. Section IV will extend the method
of obtaining the generator realizations, described in Section III, to the
co-products realized on complex
functions depending on two arguments. Finally some conclusions are given
in Section V.

\vspace{1cm}

\noindent
{\Large \bf II.- Coherent states and deforming mappings}

Let $G_{+}, G_{-}$ and $G_{0}$ be the generic expressions for the generators
of the three-dimensional algebras
 ${\cal G} = wh$, $su(2)$ and $su(1,1)$. The
(unnormalized, see below) coherent states (see
\cite{16,17,18,19,20} to which we refer for details on the general
group definition of coherent states) can be defined generically as:

$$
\vert \zeta ) = e^{\bar{\zeta} G_{+}} \vert \phi > \; , \; (\zeta \vert
= < \phi \vert e^{\zeta G_{-}} ,
\eqno{(1)}
$$

\noindent
where $\vert \phi >$ is the corresponding lowest weight state of the different
algebras,
i.e. $\vert \phi > \equiv \vert n = 0 >$ is the Fock vacuum state for the
oscillator; $\vert \phi >$ is given by $ \vert j, m = -j >$ for $su(2)$ with
$j = 1/2, 1, 3/2, ...$ and $\vert \phi > \equiv \vert k, l = 0 >$ for
$su(1,1)$ with $k = 1, \frac{3}{2}, 2, \frac{5}{2}, ...$. The round ket
indicates that the CS are unnormalized; the normalized ones are given by
$\vert \zeta > = \frac{1}{\sqrt{(\zeta \vert \zeta)}} \vert \zeta )$.

The complex variables $\zeta$ ($\zeta = \alpha, z$,  $\xi$)
 label the CS;  $\zeta$ and $\bar{\zeta}$ are the projective coordinates of
the respective coset spaces $G/H$ where $H$ is the isotropy group
of the vacuum state $\vert \phi >$, namely
$G/H = WH/U(1) \approx R^{2}, \; SU(2)/U(1) \approx S^{2}; SU(1,1)/U(1)
\approx S^{1,1}$, i.e. the two-dimensional plane, sphere and hyperboloid
respectively. Moreover $G_ {+}$ stands for the generic creation operator which
together with the two other generators $G_{-}, G_{0}$,
close into the respective Lie algebras {\cal G}.

One of the generic relations that will be extensively used below
reads

$$
[ G_{\pm}, f (G_{0})] = (f (G_{0} \mp 1) - f (G_{0})) G_{\pm}
\eqno{(2)}
$$

\noindent
for any analytic function $f$ of $G_{0}$; this relation is common
to the three algebras considered and is a consequence of the first
commutator in eqs. (4),(5) and (6) below.
 It also follows from the
definition (1) that $G_{+} \vert \zeta ) = \partial_{\bar{\zeta}} \vert
\zeta), \; (\zeta \vert G_{-} = \partial_{\zeta} (\zeta \vert$.

Let us now turn to the deforming mapping by which the generators of the
quantum $q$-oscillator, $su_{q}(2)$ and $su_{q}(1,1)$ are written uniquely in
terms
of their non-deformed counterparts.
Generically (see eg. \cite{21,22,23,24,11}),

$$
G^{q}_{\pm} = G_{\pm} F^{\pm} (G_{0})
\eqno{(3a)}
$$

$$
G^{q}_{0} = G_{0}
\eqno{(3b)}
$$

\noindent
where $G^{q}_{0} \equiv N, J_{3}, K_{3}$ and $G^{q}_{\pm} \equiv a^{\pm}_{q},
\; J^{q}_{\pm} \; ,
K^{q}_{\pm}$ and $F^{\pm} (G_{0})$ is given for each algebra by

\vspace{0.5cm}

\begin{table}[h]
\begin{center}
\begin{math}
\begin{array}{c|ccc}
 & wh_{q} & su_{q}(2) & su_{q}(1,1)\\
\hline
F^{+}(G_{0}) & \sqrt{\frac{[N+1]}{N+1}} & \sqrt{\frac{[J_{3}+j+1][J_{3}-j]}
{(J_{3}+j+1)(J_{3}-j)}} & \sqrt{\frac{[K_{3}-k+1][K_{3}+k]}{(K_{3}-k+1)
(k_{3}+k)}}\\

F^{-} (G_{0}) & \sqrt{\frac{[N]}{N}} &
\sqrt{\frac{[J_{3}-j-1][J_{3}+j]}{(J_{3}-j-1)
(J_{3}+j)}} & \sqrt{\frac{[K_{3}-k][K_{3}+k-1]}{(K_{3}-k)(K_{3}+k-1)}}
\end{array}
\end{math}
\end{center}
\end{table}

\begin{center}
Table 1
\end{center}
\vspace{0.5cm}

The deformed generators have the respective commutation relations:

$$
[N, a^{\pm}_{q}] = \pm a^{\pm}_{q} \quad , \quad [a^{-}_{q}, a^{+}_{q}] =
[N+1] - [N] \quad (q-oscillator) ;
\eqno{(4)}
$$

$$
[J^{q}_{3}, J^{q}_{\pm}] = \pm J^{q}_{\pm} \quad , \quad [J^{q}_{+}, J^{q}_{-}]
=
[2 J^{q}_{3}]) \quad (su_{q}(2)) \; ;
\eqno{(5)}
$$

\noindent
and

$$
[K^{q}_{3}, K^{q}_{\pm}] = \pm K^{q}_{\pm}\quad , \quad [K^{q}_{+}, K^{q}_{-}]
= - [2 K^{q}_{3}] \quad (su_{q} (1,1)) \; ,
\eqno{(6)}
$$

\noindent
where the square bracket is defined by $[x] \equiv \frac{q^{x} - q^{- x}}{q -
q^{-1}}$.

\vspace{1cm}

{\Large \bf III.- Realizations of the $q$-algebra generators}

We shall now develop a method which will enable us to specify complex
analytic realizations of the generators of the above quantum algebras.
This method will utilize the factorization provided by the deforming
mappings (3) by which a deformed generator is given by a non-deformed one
times a deformation operator factor $F$ (Table 1).

Let us consider the action of $G^{q}_{i}$
on a generic state $\vert \Psi >$ leading to another
state $\vert \Phi_{i}>,$

$$
G^{q}_{i} \vert \Psi > = \vert \Phi_{i} > \quad , \quad (i = +, -, 0) \quad .
\eqno{(7)}
$$

\noindent
Then,

$$
(\zeta \vert G^{q}_{i} \vert \Psi > = ( \zeta \vert \Phi_{i} >
\eqno{(8)}
$$

\noindent
may be used to define the representative $\pi_{\xi} (G^{q}_{i})$ of the
generators $G^{q}_{i}$ acting on
functions $\Psi (\zeta)$ defined on the general $\zeta$-Bargmann space,

$$
\pi_{\zeta} (G^{q}_{i}) \Psi (\zeta)\equiv \Phi_{i} (\zeta)
\eqno{(9)}
$$

\noindent
with $\Psi (\zeta) = (\zeta \vert \Psi >$ and $\Phi_{i} (\zeta) =
(\zeta \vert \Phi_{i}> .$

Since $G^{q}_{0} = G_{0}$ (eq. (3b)), in what follows
we shall concentrate on $G^{q}_{\pm}$ only and give the result for
$\pi_{\zeta} (G^{q}_{0})$ at the end.
Using eq.(3) the l.h.s. of eq.(8) is written

$$
(\zeta \vert G^{q}_{\pm} \vert \Psi > = (\zeta \vert G_{\pm}
F^{\pm} (G_{0}) \vert \Psi > = \tau_{\pm} (\zeta \vert f^{\pm}
(G_{0}) \vert \Psi > \quad ,
\eqno{(10)}
$$

\noindent
where the actions $(\zeta \vert G_{\pm}$ have been evaluated using the
following formulae for the  $wh$, $su(2)$ and $su(1,1)$ coherent states:

$$
 (\alpha \vert a^{+} = \alpha (\alpha \vert  \quad and \quad
(\alpha \vert a = \partial_{\alpha} (\alpha \vert \quad
(ordinary \quad oscillator) \; ;
\eqno{(11)}
$$

$$
(z \vert J_{+} = (z \vert (j-J_{3})z \quad and \quad
(z \vert J_{-} = (j+J_{3}) z^{-1} \quad (su(2)) \; ;
\eqno{(12)}
$$

\noindent
and

$$
(\xi \vert K_{+} = (\xi \vert (K_{3} + k) \xi \quad and \quad
(\xi \vert K_{-} = (\xi \vert (K_{3} - k) \xi^{-1} \quad (su(1,1)) \quad ,
\eqno{(13)}
$$

\noindent
and $\tau_{+} \; (\tau_{-})$ are given by $\alpha, z, \xi \;
(\partial_{\alpha},
 z^{-1}, \xi^{-1})$.
Then, by virtue of the above relations, the factorizations (3a) and Table 1, we
obtain for the $f^{\pm} (G_{0})$ defined by (10) the following values:

\newpage

\begin{table}[h]
\begin{math}
\begin{array}{c|ccc}
 & wh_{q} & su_{q}(2) & su_{q}(1,1)\\
\hline
\tau_{+} & \alpha & z & \xi\\
\tau_{-} & \partial_{\alpha} & z^{-1} & \xi^{-1}\\
f^{+} (G_{0}) & \sqrt{\frac{[N+1]}{N+1}} & (j-J_{3}) \sqrt{\frac{[J_{3}
+j+1][J_{3}-j]}{(J_{3}+j+1)(J_{3}-j)}} & (K_{3}+k) \sqrt{\frac{[K_{3}-k+1]
[K_{3}+k]}{(K_{3}-k+1)(K_{3}+k)}}\\
f^{-} (G_{0}) & \sqrt{\frac{[N]}{N}} & (j+J_{3}) \sqrt{\frac{[J_{3}+j]
[J_{3}-j-1]}{(J_{3}+j)(J_{3}-j-1)}} & (K_{3}-k) \sqrt{\frac{[K_{3}-k][K_{3}
+ k-1]}{(K_{3}-k)(K_{3}+k-1)}}
\end{array}
\end{math}
\end{table}

\begin{center}
Table 2
\end{center}

\noindent
where we have also introduced the definitions of $\tau_{\pm}$.

Using the CS defined by (1), the r.h.s of eq.(10) can be cast in the following
form:

$$
\tau_{\pm} (\zeta \vert f^{\pm} (G_{0}) \vert \Psi > = \tau_{\pm}
< \phi \vert e^{\zeta G_{-}} f^{\pm} (G_{0}) e^{-\zeta G_{-}} \, . e^{\zeta
G_{-}}
\vert \Psi >
$$
$$
= \tau_{\pm} < \phi \vert \{ f^{\pm} (G_{0}) + \zeta [G_{-}, f^{\pm}
(G_{0})] + \frac{\zeta^{2}}{2 !} [G_{-}, [G_{-}, f^{\pm} (G_{0})]] +
... \} e^{\zeta G_{-}} \vert \Psi > \; .
\eqno{(14)}
$$

\noindent
Using now relation (2) we may arrange the nested commutators in the last
equation in such a way that the functions of the generators $G_{0}$ always
appear at
the left of the monomials of $G_{-}$. This yields

$$
\tau_{\pm} (\zeta \vert f^{\pm} (G_{0}) \vert \Psi > =
\tau_{\pm} < \phi \vert \{ B^{\pm}_{0} + \zeta B^{\pm}_{1} G_{-} +
\frac{\zeta^{2}}{2 !} B^{\pm}_{2} G^{2}_{-} + ... \} e^{\zeta G_{-}}
\vert \Psi >
$$
$$
= \tau_{\pm} < \phi \vert \{ \sum_{m} \frac{\zeta^{m}}{m!} B^{\pm}_{m}
G^{m}_{-}
\} e^{\zeta G_{-}} \vert \Psi > \; ,
\eqno{(15)}
$$

\noindent where

$$
B^{\pm}_{m} \equiv (^{m}_{m}) f^{\pm} (G_{0} + m) - (^{m}_{m-1}) f^{\pm}
(G_{0} + m-1) + ... (-1)^{m} (^{m}_ {0}) f^{\pm} (G_{0})
$$
$$
=\sum_{p=0}^{m} (-1)^{m-p} (^{m}_{p}) f^{\pm}(G_{0}+p)
\eqno{(16)}
$$

\noindent
and $m = 0,1,2,...$

As the $G_{0}$ generators are all diagonal
in the basis of the states constructed
from their lowest weight $\vert \phi >$, we will have generically that

$$
< \phi \vert B^{\pm}_{m} = < \phi \vert b^{\pm}_{m} \quad ,
\eqno{(17)}
$$

\noindent
where the numerical eigenvalues $b^{\pm}_{m}$ will be evaluated explicitly
below
for each of the three algebras considered. Using (17), (1) and that
$(\zeta\vert G_{-}=\partial_{\zeta}(\zeta\vert$, eq. (15) now
leads to

$$
\tau_ {\pm} (\zeta \vert f^{\pm} (G_{0}) \vert \Psi > = \tau_{\pm}
\{ b^{\pm}_{0} + b^{\pm}_{1} \zeta \partial_{\zeta} + \frac{b^{\pm}_{2}}
{2!} \zeta^{2} \partial^{2}_{\zeta} + ... \} (\zeta \vert \Psi >
\equiv \pi_{\zeta} (G^{q}_{\pm}) \Psi (\zeta) ,
\eqno{(18)}
$$

\noindent
where $\pi_{\xi} (G^{q}_{\pm})$ defines the realization of $G^{q}_{\pm}$ on the
functions $\Psi (\zeta)$. Let us notice that in the $su_{q}(2)$ case, the
expansions (15) and (18)
terminate since then $G^{2j+1}_{\pm} \equiv J^{2j+1}_{\pm}$ is zero on
any vector of the representation space.

Collecting now all the above results and replacing $\tau_{\pm}$
by their values (Table 2) the following realizations
for the deformed generators are obtained:

\vspace{1cm}

\noindent
{\it a) Quantum deformed oscillator}

$$
\pi_{\alpha} (a^{+}_{q}) = \partial_{\alpha} \sum^{\infty}_{m=0}
\frac{b^{+}_{m}}
{m!} \alpha^{m} \partial^{m}_{\alpha}
\eqno{(19a)}
$$

$$
\pi_{\alpha} (a^{-}_{q}) = \alpha \sum^{\infty}_{m=0} \frac{b^{-}_{m}}{m!}
\alpha^{m} \partial^{m}_{\alpha}
\eqno{(19b)}
$$

$$
\pi_{\alpha} (N) = \alpha \partial_{\alpha} \quad  ,
\eqno{(19c)}
$$

\noindent
where

$$
b^{\pm}_{m} = (^{m}_{m}) f^{\pm}_{m} - (^{m}_{m-1}) f^{\pm}_{m-1} + ...
(-1)^{m}
(^{m}_{0}) f^{\pm}_{0}=\sum_{p=0}^{m} (-1)^{m-p}(^{m}_{p})f^{\pm}_{p}
\eqno{(20a)}
$$

\noindent
with

$$
f^{+}_{p} = \sqrt{\frac{[p+1]}{p+1}} \quad and \quad f^{-}_{p} =
\sqrt{\frac{[p]}{p}}
\eqno{(20b)}
$$
since for instance $<0\vert f^{+}(N+p)=<0\vert\sqrt{\frac{[p+1]}{p+1}}\equiv
<0\vert f^{+}_{p}$.

The carrier space for this realization $L^{2} (C, \frac{1}{\pi} \frac{d^{2}
\alpha}
{(\alpha \vert \alpha)})$ possesses an orthonormal basis formed by the
monomials $\{ \gamma_{n} (\alpha) \equiv (\alpha \vert n > = \frac{\alpha^{n}}
{\sqrt{n!}} ; n=0,1,2,... \}$, where

$$
< \gamma_{n} (\alpha), \gamma_{n'}(\alpha) > = \delta_{n n'} =
\frac{1}{\pi} \int d^{2} \alpha e^{- \alpha \bar{\alpha}}
\bar{\gamma}_{n} (\alpha) \gamma_{n'} (\alpha)
\eqno{(21)}
$$

\noindent
since

$$
{\bf 1} = \frac{1}{\pi} \int d^{2} \alpha e^{\alpha \bar{\alpha}} \vert \alpha)
(\alpha \vert \quad .
\eqno{(22)}
$$

The action $\pi_{\alpha}$ of the generators $a^{\pm}_{q},\;N$ on the basis
vectors is given by the usual expressions

$$
\pi_{\alpha} (a^{+}_{q})\; \frac{\alpha^{n}}{\sqrt{n!}} = \sqrt{[n+1]}
\,\frac{\alpha^{n+1}}{\sqrt{(n+1)!}}
\eqno{(23a)}
$$

$$
\pi_{\alpha} (a^{-}_{q})\; \frac{\alpha^{n}}{\sqrt{n!}} = \sqrt{[n]}\;
\frac{\alpha^{n-1}}{\sqrt{(n-1)!}}
\eqno{(23b)}
$$

$$
\pi_{\alpha} (N)\; \frac{\alpha^{n}}{\sqrt{n!}} = n
\;\frac{\alpha^{n}}{\sqrt{n!}} \quad .
\eqno{(23c)}
$$

\noindent
In the $q \rightarrow 1$ limit
eq. (20b) shows that $f^{\pm}_{p}\rightarrow 1$ and it is simple
to check in eq. (20a) that
 $b^{\pm}_{0} \rightarrow 1$ and
$b^{\pm}_{i \neq 0} \rightarrow 0$. As a result, the previous expressions
yield the expected 'classical' limit relations
 $\pi_{\alpha} (a^{+}_{q}) \rightarrow \alpha$,
$\pi_{\alpha} (a^{-}_{q})\rightarrow \partial_{\alpha}$
(and of course $\pi_{\alpha}(N)=\alpha\partial_{\alpha}$)
of the oscillator Bargmann algebra.

\vspace{0.5cm}

\noindent
{\it b)}  $su_{q}(2)$ {\it algebra}

Proceeding similarly, we find

$$
\pi_{z} (J^{q}_{+}) = z \sum^{2j}_{m=0} \frac{b^{+}_{m}}{m!} z^{m}
\partial^{m}_{z}
\eqno{(24a)}
$$

$$
\pi_{z} (J^{q}_{-}) = z^{-1} \sum^{2j}_{m=0} \frac{b^{-}_{m}}{m!} z^{m}
\partial^{m}_{z}
\eqno{(24b)}
$$

$$
\pi_{z} (J^{q}_{3}) = -j + z \partial_{z} \quad ,
\eqno{(24c)}
$$

\noindent
where now

$$
b^{\pm}_{m} = \sum^{m}_{p=0} (-1)^{m-p} (^{m}_{p}) f^{\pm}_{p}
\eqno{(25a)}
$$

\noindent
and

$$
f^{+}_{p} = (2j-p) \sqrt{\frac{[2j-p][p+1]}{(2j-p)(p+1)}} \quad , \quad
f^{-}_{p} = p \sqrt{\frac{[p][p-2j-1]}{p (p-2j-1)}}
\eqno{(25b)}
$$
which follow from computing $<j,-j\vert f^{\pm}(J_{3}+p)$ using Table 2.

The Hilbert representation space
$L^{2} (C - \{ \infty \}, \frac{(2j+1)}{\pi} \frac{d^{2} z}{(1 + \vert z
\vert^{2})^{2j+2}})$
admits the following basis $\{ \gamma_{n} (z) \equiv ({z} \vert n > =
(^{2j}_{j+n})^{1/2} z^{j+n}, n=-j,...,j \}$ with orthonormality conditions,

$$
< \gamma_{n} (z), \gamma_{n'} (z) > = \delta_{n n'} = \frac{(2j+1)}{\pi}
\int \frac{d^{2} z}{(1+\vert z \vert^{2})^{2j+2}} \bar{\gamma}_{n}
(z) \gamma_{n'} (z) \quad ,
\eqno{(26)}
$$

\noindent
where the new measure is designed so that the $\gamma_{n} (z)$ states are
orthonormal since

$$
{\bf 1} = \frac{(2j + 1)}{\pi} \int \frac{d^{2}z }{(1+\vert 2 \vert^{2})^{2j
+2}} \vert z) (z \vert \quad .
\eqno{(27)}
$$

The action of the deformed generator on the unit vector of the representation
space reproduces the familiar $q$-Fock expressions

$$
\pi_{z} (J^{q}_{+})\; \gamma_{m}(z) = \sqrt{[j-m][j+m+1]}\; \gamma_{m+1}(z)
\eqno{(28a)}
$$

$$
\pi_{z} (J^{q}_{-})\;\gamma_{m}(z) = \sqrt{[j+m][j-m+1]}\;
\gamma_{m-1}(z)
\eqno{(28b)}
$$

\noindent
and

$$
\pi_{z} (J^{q}_{3})\;\gamma_{m}(z) = m \gamma_{m}(z) \quad .
\eqno{(28c)}
$$

In the zero deformation limit, $q \rightarrow 1$,
$f^{+}_{p}\rightarrow (2j-p)$ and $f^{-}_{p}\rightarrow p$ (eq. (25b)). Then,
 $b^{+}_{0} \rightarrow
2j$, $b^{+}_{1} \rightarrow -1$ and all the other $b^{+'}s$ become zero.
Thus, $\pi_{z} (J^{q}_{+}) \rightarrow 2jz - z^{2} \partial_{z}$. Similarly
$b^{-}_{1} \rightarrow 1$ and the other $b^{-}$'s are zero
in the same limit and we obtain $\pi_{z} (J^{q}_{-}) \rightarrow \partial_{z}$,
the realization $\pi_{z}(J_{3})=-j+z\partial_{z}$ remaining unaffected.
Thus, the reduction to the ordinary coset representation (see e.g.
\cite{17,18,19}) of the
su(2) algebra is provided by the classical $q \rightarrow 1$ limit.

\vspace{0.5cm}

\noindent
{\it c)} $su_{q} (1,1)$ {\it algebra}

Finally, for the $su_{q}(1,1)$ generators we find the realization

$$
\pi_{\xi} (K^{+}_{q}) = \xi \sum^{\infty}_{m=0} \frac{b^{+}_{m}}{m!}
\xi^{m} \partial^{m}_{\xi}
\eqno{(29a)}
$$

$$
\pi_{\xi} (K^{-}_{q}) = \xi^{-1} \sum^{\infty}_{m=0} \frac{b^{-}_{m}}{m!}
\xi^{m} \partial^{m}_{\xi}
\eqno{(29b)}
$$

$$
\pi_{\xi} (K^{3}_{q}) = k + \xi \partial_{\xi} \quad ,
\eqno{(29c)}
$$

\noindent
with
$$
b^{\pm}_{m} = \sum^{m}_{p=0} (-1)^{m-p} (^{m}_{p}) f^{\pm}_{p} \quad,
\eqno{(30a)}
$$

\noindent
where now Table 2 gives for  $<k,0\vert f^{\pm}_{p} (K_{3}+p)$ the
expressions

$$
f^{+}_{p} = (2k + p) \sqrt{\frac{[p+1][2k+p]}{(p+1)(2k+p)}} \quad
and \quad f^{-}_{p} = p \sqrt{\frac{[p][2k+p-1]}{p(2k+p-1)}} \; .
\eqno{(30b)}
$$

The representation space is now the Hilbert space of square integrable
functions with support on the open unit disk on the complex plane,
$D = \{ \xi \in C : \vert \xi \vert^{2} < 1 \}$, denoted by
$L^{2} (D, \frac{(2k-1)}{\pi} \frac{d^{2} \xi}{(1- \vert \xi \vert^{2})
^{- (2k-2)}})$
which has the basis of vectors
$\{ \gamma_{n} (\xi) \equiv (\xi \vert n > = \left( \frac{\Gamma
(2 k+n)}{n! \Gamma (2k)} \right)^{\frac{1}{2}} \xi^{n}, n= 0,1,2,... \}$,
satisfying the orthonormality condition

$$
< \gamma_{n} (\xi), \gamma_{n'} (\xi) > = \delta_{nn'} = \frac{(2k-1)}
{\pi} \int_{\vert \xi \vert <1} \frac{d^{2} \xi}{(1- \vert \xi \vert^{2})^{ -
(2 k-2)}} \bar{\gamma}_{n} (\xi) \gamma_{n'} (\xi) \quad \eqno{(31a)}
$$
\noindent
corresponding to the completeness relation

$$
{\bf 1}= \frac{(2k-1)}
{\pi} \int_{\vert \xi \vert <1} \frac{d^{2} \xi}{(1- \vert \xi \vert^{2})^{ -
(2 k-2)}} \vert\xi) (\xi\vert \quad .
\eqno{(31b)}
$$

\noindent
The action on the above basis monomials is given by

$$
\pi_{\xi} (K^{q}_{+}) \;\gamma_{n} (\xi) = \sqrt{[2k+n][n+1]}\;
\gamma_{n+1} (\xi) \quad ,
\eqno{(32a)}
$$

$$
\pi_{\xi} (K^{q}_{-})\; \gamma_{n} (\xi) = \sqrt{[2k+n+1][n]}
\;\gamma_{n-1} (\xi) \quad ,
\eqno{(32b)}
$$

\noindent
and

$$
\pi_{\xi} (K^{q}_{3}) \gamma_{n} (\xi) = (k+n) \gamma_{n} (\xi) \quad .
\eqno{(32c)}
$$

If there is no deformation $f^{+}_{p}\rightarrow 2k+p$ and $f^{-}_{p}
\rightarrow p$ (eq. (30b)). Then, eq. (30a) gives
$b^{-}_{n=0} = b^{-}_{n \geq 2} =0$ and $b^{-}_{1} =1$ so $\pi_{\xi}
(K^{q}_{-})
\rightarrow \pi_{\xi} (K_{-}) = \partial_{\xi}$, and $\pi_{\xi} (K^{q}_{+})
\rightarrow \pi_{\xi} (K_{+}) = 2 k \xi + \xi ^{2} \partial_{\xi}$;
$\pi_{\xi}(K_{3})$ is again given by (32c).

Having derived a complex analytic realization of the three algebra generators
we shall give in the next Section the realizations of the corresponding
co-products.

\vspace{1cm}

{\Large \bf IV.- Analytic realizations for the co-products}

In order to construct a realization of the co-products we shall write them
generically in the form

$$
\Delta G^{q}_{\pm} = G^{q}_{\pm} \otimes g (G_{0}) + g' (G_{0}) \otimes
G^{q}_{\pm}
\eqno{(33a)}
$$

$$
\Delta G^{q}_{0} = G_{0} \otimes 1 + 1 \otimes G_{0} \quad ,
\eqno{(33b)}
$$

\noindent
where $g(G_{0})$ and $g' (G_{0})$ will be explicitly found for each algebra
separately.

As the co-product of each algebra generator acts in the tensor product
of the representation space of the algebra, we shall look for a
analytic functional realization carried by functions of two variables.
To this aim consider

$$
(\zeta_{1} \vert \otimes (\zeta_{2} \vert \Delta G^{q}_{\pm}
\vert \Psi > = ( \zeta_{1} \vert \otimes ( \zeta_{2} \vert
G^{q}_{\pm} \otimes g (G^{q}_{0}) \vert \Psi > +
$$
$$
+ (\zeta_{1} \vert \otimes (\zeta_{2} \vert g' (G^{q}_{0})
\otimes G^{q}_{\pm} \vert \Psi > \quad .
\eqno{(34)}
$$

\noindent
Using the deformation mapping (3) and eq. (10) this equation gives

$$
(\zeta_{1} \vert \otimes (\zeta_{2} \vert \Delta G^{q}_{\pm} \vert
\Psi > = \tau^{1}_{\pm}  (\zeta_{1} \vert \otimes (\zeta_{2}
\vert f^{\pm} (G_{0}) \otimes g (G_{0}) \vert \Psi > +
$$

$$
+ \tau^{2}_{\pm} (\zeta_{1} \vert \otimes (\zeta_{2} \vert
g' (G_{0}) \otimes f^{\pm} (G_{0}) \vert \Psi > \equiv I_{1} + I_{2} \quad ,
\eqno{(35)}
$$

\noindent
where $\tau^{1}_{\pm}$ and $\tau^{2}_{\pm}$, where the
indices refer to the first or second
term in the tensor product, are the same as in Table 2
for the different algebras. We shall now compute $I_{1}$ and limit
ourselves to give the formulae for $I_{2}$, since they are derived in
exactly the same fashion. Using again definition (1) we find

$$
I_{1} = \tau^{1}_{\pm} < \phi \vert \otimes < \phi \vert (e^{\zeta_{1} G_{-}}
\otimes e^{\zeta_{2} G_{-}}) (f^{\pm} (G_{0}) \otimes g (G_{0}))
(e^{-\zeta_{1} G_{-}} \otimes e^{- \zeta_{2} G_{-}})
$$
$$
(e^{\zeta_{1} G_{-}} \otimes e^{\zeta_{2} G_{-}}) \vert \Psi > =
$$
$$
= \tau^{1}_{\pm} < \phi \vert \otimes < \phi \vert (e^{\zeta_{1} G_{-}} f^{\pm}
(G_{0}) e^{- \zeta_{1} G_{-}} \otimes e^{\zeta_{2} G_{-}} g (G_{0})
e^{- \zeta_{2} G_{-}})
$$
$$
(e^{\zeta_{1} G_{-}} \otimes e^{\zeta_{2} G_{-}}) \vert \Psi > \quad .
\eqno{(36)}
$$

\noindent
Expanding the exponentials in the above formula and rearranging the
ensuing nested commutators for $f^{\pm}(G_{0})$ and $g(G_{0})$ as we did
for eqs. (14,15) we obtain

$$
I_{1} = \tau^{\pm}_{1} < \phi \vert \otimes < \phi \vert \left( \sum^{l}_{n=0}
\frac{\zeta^{n}_{1}}{n!} B^{\pm}_{n} G^{n}_{-} \otimes \sum^{l}_{m=0}
\frac{\zeta^{m}_{2}}{m!} C_{m} G^{m}_{-} \right)  \left( e^{\zeta_{1} G_{-}}
\otimes
e^{\zeta_{2} G_{-}} \right) \vert \Psi > \; ,
\eqno{(37)}
$$

\noindent
where $l$ is the appropriate limit for each algebra, $B^{\pm}_{n}$ are the
functions of $G_{0}$ introduced in (15),(16) and (17), and the
operators $C_{m}$ are defined (cf. (16)) by the expansion

$$
C_{m} = (^{m}_{m}) g (G_{0} + m) - (^{m}_{m-1}) g (G_{0} + m-1) + ...+
(-1)^{m} (^{m}_{0}) g (G_{0})
$$
$$=\sum_{p=0}^{m} (-1)^{m-p} (^{m}_{p}) g(G_{0}+p)\quad .
\eqno{(38)}
$$

\noindent
On the vacuum,
$< \phi \vert C_{m} = < \phi \vert c_{m} $ (the numerical eigenvalues
will be specified later for each algebra separately). Then,  using that
$( \zeta \vert G_{-} = \partial_{\zeta} (\zeta \vert$, eq.(37)
may be written as

$$
I_{1} = \tau^{\pm}_{1} (\zeta_{1} \vert \otimes (\zeta_{2}
\vert \left[ \sum^{l}_{n,m=0} \frac{b^{\pm}_{n}}{n!} \frac{c_{m}}{m!}
\zeta^{n}_{1} \zeta^{m}_{2} G^{n}_{-} \otimes G^{m}_{-} \right]
\vert \Psi >
$$
$$
= \tau^{\pm}_{1} \left[ \sum^{l}_{n,m=0} \frac{b^{\pm}_{n}}{n!} \frac{c_{m}}
{m!} \zeta^{n}_{1} \zeta^{m}_{2} \partial^{n}_{\zeta_{1}}
\partial^{m}_{\zeta_{2}}
\right] (\zeta_{1} \vert \otimes (\zeta_{2} \vert ) \vert
\Psi >
$$
$$
= \pi_{\zeta_{1}} (G^{q}_{\pm}) \pi_{\zeta_{2}} (g (G_{0})) \Psi (\zeta_{1},
\zeta_{2}) \; ,
\eqno{(39)}
$$

\noindent
where we have used (18) for the identification

$$
\pi_{\zeta_{1}} (G^{q}_{\pm}) = \tau^{\pm}_{1} \sum^{l}_{n=0}
\frac{b^{\pm}_{n}}
{n!} \zeta^{n}_{1} \partial^{n}_{\zeta 1}
\eqno{(40a)}
$$

\noindent
and written

$$
\pi_{\zeta_{1}} (g (G_{0})) = \sum^{l}_{m=0} \frac{c_{m}}{m!}
\zeta^{m}_{2} \partial^{m}_{\zeta_{2}} \quad .
\eqno{(40b)}
$$

\noindent
Collecting all results for eq.(33) the realization of the co-product is
obtained,

$$
\pi_{\zeta_{1} \zeta_{2}} (\Delta G^{q}_{\pm}) \Psi (\zeta_{1}, \zeta_{2})
\equiv (\zeta_{1} \vert \otimes (\zeta_{2} \vert
\Delta G^{q}_{\pm} \vert \Psi > =
$$

$$
= \left[ \pi_{\zeta_{1}} (G^{q}_{\pm}) \pi_{\zeta_{2}} (g (G_{0})) +
\pi_{\zeta_{1}}
(g' (G_{0})) \pi_{\zeta_{2}} (G^{q}_{\pm}) \right]
\Psi (\zeta_{1}, \zeta_{2}) \; ,
\eqno{(41)}
$$

\noindent
and, similarly,

$$
\pi_{\zeta_{1} \zeta_{2}} (\Delta G^{q}_{0}) = \left[ \pi_{\zeta_{1}} (G_{0}) +
\pi_{\zeta_{2}} (G_{0}) \right] \Psi (\zeta_{1}, \zeta_{2}) \quad .
\eqno{(42)}
$$

As there is no satisfactory co-product and bialgebra structure for the
$q$-oscillator, we shall now specialize to the $su_{q}(2)$ and $su_{q}(1,1)$
cases for
which we have, respectively,

$$
\Delta J^{q}_{\pm} = J^{q}_{\pm} \otimes q^{J^{q}_{3}} + q^{-J^{q}_{3}}
\otimes J^{q}_{\pm}
\eqno{(43a)}
$$

$$
\Delta J^{q}_{3} = J^{q}_{3} \otimes 1 \otimes 1 \otimes J^{q}_{3} \quad ,
\eqno{(43b)}
$$

\noindent
and

$$
\Delta K^{q}_{\pm} = K^{q}_{\pm} \otimes q^{K^{q}_{3}} + q^{- K^{q}_{3}}
\otimes K^{q}_{\pm},
\eqno{(44a})
$$

$$
\Delta K^{q}_{3} = K^{q}_{3} \otimes 1 + 1 \otimes K^{q}_{3} \quad .
\eqno{(44b)}
$$
\noindent
so that $g(G_{0})=q^{G_{o}}$ and $g'(G_{0})=q^{-G_{0}}$ with $G_{0}=J_{3},
K_{3}$ respectively.
\noindent

a) $su_{q} (2)$

For the $su_{q}(2)$ case eq. (41) and (42) give

$$
\pi_{z_{1} z_{2}} (\Delta J^{q}_{\pm}) = \pi_{z_{1}} (J^{q}_{\pm}) \pi_{z_{2}}
(q^{J^{q}_{3}}) + \pi_{z_{1}} (q^{- J^{q}_{3}}) \pi_{z_{2}} (J^{q_{\pm}})
\eqno{(45a)}
$$

$$
\pi_{z_{1} z_{2}} (\Delta J^{q}_{3}) = \pi_{z_{1}} (J^{q}_{3}) + \pi_{z_{2}}
(J^{q}_{3})
\eqno{(45b)}
$$

\noindent
where (cf. eqs. (40))

$$
\pi_{z_{1,2}} (J^{q}_{\pm}) = z^{\pm 1}_{1,2} \sum^{2j}_{n=0}
\frac{b^{\pm}_{n}}
{n!} z^{n}_{1,2} \partial^{n}_{z_{1,2}}
\eqno{(46a)}
$$

$$
\pi_{z_{1,2}} (J^{q}_{3}) = -j + z_{1,2} \partial_{z_{1,2}}
\eqno{(46b)}
$$

$$
\pi_{z_{1,2}} (q^{\pm J^{q}_{3}}) = \sum^{2j}_{m=0} \frac{c_{m}^{\pm}}{m!}
z^{n}_{1,2} \partial^{m}_{z_{1,2}} \quad ,
\eqno{(46c)}
$$

\noindent
where the $b$'s are given by eq.(25) and (cf. eq. (38))

$$
c^{\pm}_{m} = (^{m}_{m}) q^{\pm (-j+m)} - (^{m}_{m-1}) q^{\pm(-j+m-1)} + ...
(-1)^{m} q^{\mp j}
=\sum^{m}_{p=0} (-1)^{m-p} (^{m}_{p}) q^{\pm(-j+p)} \; .
\eqno{(47)}
$$

The realization of the co-product acts on the tensor product of two
representation
spaces which is spanned by the basis

$$
\gamma_{n m} (z_{1}, z_{2}) \equiv ({z}_{1} \vert \otimes ({z}_{2}
\vert . \vert n > \otimes \vert m > = (^{2j}_{j+n})^{\frac{1}{2}} (^{2j}_{j+m})
^{\frac{1}{2}} z^{j+n}_{1} z^{j+m}_{2},
\eqno{(48)}
$$

\noindent
and which satisfies an orthonormality relation induced by that in each of the
factors (eq. (26)). On the basis elements $\gamma_{n m} (z_{1}, z_{2})$ the
action of $\pi_{z_{1} z_{2}}$ is given by (cf. eq.(28)),

$$
\pi_{z_{1} z_{2}} (\Delta J^{q}_{\pm}) \gamma_{n,m} (z_{1}, z_{2}) =
$$

$$
= \sqrt{[j \mp  n][j \pm  n+1]} q^{m} \gamma_{n \pm 1, m} + q^{-n}
\sqrt{[j\mp m][j \pm m +1]} \gamma_{n, m \pm 1} \; ,
\eqno{(49a)}
$$

$$
\pi_{z_{1} z_{2}} (\Delta J^{q}_{3}) \gamma_{n,m} (z_{1}, z_{2}) =
(n+m) \gamma_{n,m} (z_{1}, z_{2}) \quad .
\eqno{(49b)}
$$

\noindent{\it b)} $su_{q} (1,1)$

Finally, for the $su_{q}(1,1)$ case, eq. (41), we find

$$
\pi_{\xi_{1} \xi_{2}} (\Delta K^{q}_{\pm}) = \pi_{\xi_{1}} (K^{q}_{\pm})
\pi_{\xi_{2}} (q^{K^{q}_{3}}) + \pi_{\xi_{1}} (q^{- K^{q}_{3}}) \pi_{\xi_{2}}
(K^{q}_{\pm})
\eqno{(50a)}
$$

$$
\pi_{\xi_{1} \xi_{2}} (\Delta K^{q}_{3}) = \pi_{\xi_{1}} (K^{q}_{3}) +
\pi_{\xi_{2}} (K^{q}_{3}) \quad ,
\eqno{(50b)}
$$

\noindent
where

$$
\pi_{\xi_{1,2}} (K^{q}_{\pm}) = \xi^{\pm 1}_{1,2} \sum^{\infty}_{m=0}
\frac{b^{\pm}_{m}}{m!} \xi^{m}_{1,2} \partial^{m}_{\xi_{1,2}}
\eqno{(51a)}
$$

$$
\pi_{\xi_{1,2}} (K^{q}_{\pm}) = k +
 \xi_{1,2} \partial_{\xi_{1,2}}
\eqno{(51b)}
$$

$$
\pi_{\xi_{1,2}} (q^{\pm K^{q}_{3}}) = \sum^{\infty}_{m=0} c^{\pm}_{m}
\xi^{m}_{1,2}
\partial^{m}_{\xi_{1,2}} \quad ,
\eqno{(51c)}
$$

\noindent
where the coefficients $b^{\pm}_{m}$ are given in eq.(28) and the $c^{\pm}_{m}$
in this case read (cf. eq. (47))

$$
c^{\pm}_{m} = (^{m}_{m}) q^{\pm (k+m)} - (^{m}_{m-1}) q^{\pm(k+m-1)} + ... +
(-1)^{m} q^{\pm k}
=\sum_{p=0}^{m}(-1)^{m-p} q^{\pm (k+p)} \quad .
\eqno{(52)}
$$

As for $su_{q}(2)$, the orthonormal basis which spans the representation
space of the above co-product realization inherits its orthonormality
properties
from eq.(31) and reads

$$
\gamma_{n m} (\xi_{1}, \xi_{2}) \equiv ({\xi}_{1} \vert \otimes ({\xi}
_{2} \vert . \vert n > \otimes \vert m > =
$$

$$
= \left( \frac{\Gamma (2k+n)}{n! \Gamma (2k)} \right)^{\frac{1}{2}}
\left( \frac{\Gamma (2k+m)}{m! \Gamma (2k)} \right)^{\frac{1}{2}}
\xi^{n}_{1} \xi^{m}_{2} \quad .
\eqno{(53)}
$$

Extending the action expressed by eqs. (32) to the co-product realization
of the $su_{q}(1,1)$ generators we obtain

$$
\pi_{\xi_{1} \xi_{2}} (\Delta K^{q}_{+}) \gamma_{n,n} (\xi_{1}, \xi_{2}) =
$$
$$
= \sqrt{[n+1][n+2k]} q^{k+n} \gamma_{n+1,m} (\xi_{1}, \xi_{2}) +q^{-(k+m)}
\sqrt{[m+1][m+2k]} \gamma_{n,m+1} (\xi_{1}, \xi_{2})
\eqno{(51a)}
$$
$$
\pi_{\xi_{1} \xi_{2}} (\Delta K^{q}_{-}) \gamma_{n,m} (\xi_{1}, \xi_{2})
= \sqrt{[n][n-1+2k]} q^{k+n} \gamma_{n-1,m} (\xi_{1}, \xi_{2}) +
$$
$$
q^{- (k+m)} \sqrt{[m][m-1+2k]} \gamma_{n,m-1} (\xi_{1}, \xi_{2})
\eqno{(54b)}
$$

\noindent
and

$$
\pi_{\xi_{1} \xi_{2}} (\Delta K^{q}_{3}) \gamma_{n,m} (\xi_{1}, \xi_{2}) =
(2k+n+m) \gamma_{n,m} (\xi_{1}, \xi_{2}) \quad .
\eqno{(54c)}
$$
\vskip 1cm
{\Large \bf V.- Conclusions}

In this paper we have looked for realizations of quantum algebras in terms
of ordinary differential operators. We have found their expression for
certain $q$-algebras for which there exist functional deforming mappings
relating the deformed and non-deformed generators which allow us to write
explicitly the deformed generators as elements of the original enveloping
algebra. The
$q$-oscillator, $su_{q}(2)$ and $su_{q}(1,1)$ quantum algebras
are particular cases of this
class, and the method of constructing realizations relies on
the ordinary coherent states for the undeformed Lie algebras.

Finally, we recall that the realization  for the
deformed algebra generators obtained here is given in terms of
a series of powers of derivatives with $q$-dependent coefficients which
in the classical $q \rightarrow 1$ limit reproduces the vector field
generators of the Lie algebra. Thus, the appearance of ordinary but higher
order
derivatives provides an alternative way of describing the deformation process.

\end{document}